\documentclass[12pt]{article}
\usepackage{caption}
\usepackage[utf8]{inputenc}
\usepackage{amsmath,amsthm,bm,amsfonts,mathrsfs,mathtools,hyperref,lipsum}
\usepackage[numbered]{bookmark}
\usepackage{mdframed}
\usepackage{courier}
\usepackage{lineno}
\usepackage{authblk}
\hypersetup{
    colorlinks,
    citecolor=black,
    filecolor=black,
    linkcolor=black,
    urlcolor=black
}
\usepackage[a4paper, total={7in, 10in}]{geometry}
\usepackage{graphicx, float, multicol, balance}
\usepackage[version=4]{mhchem}
\usepackage[sorting=none, maxnames=5, style=nature]{biblatex}

\DeclareUnicodeCharacter{2217}{*}
\title{Spin-Dependent Amyloid Self-Assembly on Magnetic Substrates}

\author[1]{Yael Kapon$^\#$}
\author[1]{Dror Merhav$^\#$}
\author[2]{Gal Finkelstein-Zuta}
\author[3]{Omer Blumen}
\author[4]{Naomi Melamed Book}
\author[5]{Yael Levi-Kalisman}
\author[5]{Ilya Torchinsky}
\author[1]{Shira Yochelis}
\author[3]{Daniel Sharon}
\author[6]{Lech Tomasz Baczewski}
\author[2]{Ehud Gazit}
\author[1]{Yossi Paltiel\thanks{paltiel@mail.huji.ac.il}}

\affil[1]{Institute of Applied Physics, The Hebrew University, Jerusalem 9190401, Israel}
\affil[2]{The Shmunis School of Biomedicine and Cancer Research, George S. Wise Faculty of Life Sciences, Tel Aviv University, Tel Aviv, Israel}
\affil[3]{Institute of Chemistry, Hebrew University of Jerusalem, Jerusalem 91904, Israel}
\affil[4]{Bio-Imaging Unit, The Alexander Silberman Institute of Life Science, The Hebrew University, Jerusalem 9190401, Israel}
\affil[5]{The Harvey M. Krueger Family Center for Nanoscience and Nanotechnology, The Hebrew University, Jerusalem 9190401, Israel}
\affil[6]{Institute of Physics, Polish Academy of Sciences, Warsaw 02668, Poland}

\date{\today}

\begin{document}
\maketitle

\renewcommand{\thefootnote}{\#}
\footnotetext{These authors contributed equally to this work.}

\begin{abstract}
Protein aggregation into insoluble amyloid-like fibrils is implicated in a wide range of diseases and understanding its nucleation process is a key for mechanistic insights and advancing therapeutics. The electronic charge of the amyloidogenic monomers significantly influences their self-assembly process. However, the impact of electron spin interactions between monomers on amyloid nucleation has not been considered yet. Here, we studied amyloid formation on magnetic substrates using Scanning Electron Microscopy (SEM), fluorescence microscopy, and Attenuated Total Reflection Fourier Transform Infrared (ATR-FTIR) Spectroscopy. We observed a preferred magnetization orientation of the ferromagnetic layer for fibril formation, leading to twice as many and significantly longer fibrils (up to 20 times) compared to the opposite magnetization orientation. This preference is related to monomer chirality. Additionally, fibril structure varied with substrate magnetization orientation. Our findings suggest a transient spin polarization in monomers during self-assembly, driven by the Chiral Induced Spin Selectivity (CISS) effect. These effects are consistent for various molecule length scales, from A${\beta}$ polypeptide to dipeptides and single amino acids, indicating a fundamental spin-based dependence on biomolecular aggregation that could be harnessed for targeted interventions for amyloid-related diseases. 
\end{abstract}

\subsection*{Introduction}
Amyloid fibrils are highly ordered protein aggregates implicated in numerous pathological conditions, including Alzheimer’s, Parkinson’s, and Type II Diabetes, where their formation disrupts normal cellular functions.\cite{knowles2014amyloid} At the same time, their highly ordered nanostructure has been explored as functional biomaterials in nanotechnology due to their unique biophysical properties.\cite{wei2017self}

Even very short aromatic peptide fragments, including diphenylalanine peptides (Phe-Phe), representing the core recognition motif within the Amyloid ${\beta}$ (A${\beta}$) polypeptide, form well-ordered nanotubular assemblies with amyloid-like structural signatures. \cite{reches2003casting} These phenylalanine residues (Phe19 and Phe20) are suggested to mediate intermolecular interactions between polypeptide chains. \cite{reches2003casting} This hypothesis is supported by their role as key components in the formation of amyloid-like fibrils. \cite{soto1998beta,tjernberg1996arrest} Remarkably, even single phenylalanine amino acids can form similar amyloid-like structures \cite{adler2012phenylalanine}, underscoring the fundamental role of aromatic interactions in self-assembly processes. \cite{zaguri2021kinetic}

The influence of external fields, such as electric and magnetic fields, on amyloid formation has been a topic of increasing interest. \cite{pandey2017modulation,wang2011charged} Due to the strong electric dipole and low diamagnetism of the monomers, these fields can modulate molecular interactions and orientation, thereby impacting the nucleation and growth of fibrils. Magnetic fields required for such effects tend to be extremely strong, and the low diamagnetism of peptides necessitates using magnetic fields exceeding 10 Tesla for significant influence. \cite{a2007alignment} In addition, using electric fields during the peptide self-assembly process can induce uniform and controlled polarization. This effect however requires also a relatively high electric field, of the order of 1kV/mm. \cite{nguyen2016self,su2023electric}

While these effects are relatively well-documented, the role of electron spin interaction remains largely unexplored in biological systems. Many biological molecules, including DNA \cite{gohler2011spin,xie2011spin}, various proteins \cite{mishra2013spin,niman2023bacterial,varade2018bacteriorhodopsin,gupta2023spin,sang2021temperature}, and sugars \cite{al2022spin,aminadav2024chiral,tassinari2019enantioseparation} demonstrated significant spin selective conductance due to the Chiral-Induced Spin Selectivity (CISS), a phenomenon relating molecular chirality to the electron spin \cite{naaman2019chiral}. While it traditionally requires a current through the chiral molecule, recent advancements demonstrated that spin polarization could also appear due to transient charge displacement during adsorption on magnetic substrates. Achieving enantio-separation due to interactions with perpendicularly magnetized substrates. \cite{banerjee2018separation,lu2021enantiospecificity,safari2024enantioselective} These spin-exchange interactions are influenced by the adsorption kinetics, with prolonged adsorption times leading to a loss of selectivity. Moreover, in a biological context, spin polarization was observed due to transient charge displacement during intermolecular interactions \cite{banerjee2020long,kumar2017chirality,wei2023examining}, opening new possibilities for exploring spin-dependent phenomena in biological systems (\autoref{Abeta}a). 

While spin polarization has been measured in short peptides and single amino acids, generally the longer the molecule and its electrical dipole, the larger its spin polarization.\cite{gohler2011spin,kettner2015spin} This makes growing systems, such as self-assembly or polymerization, particularly intriguing due to their evolving properties and potential accumulation of electric dipoles. For example, nucleation rates of conglomerate crystallization \cite{tassinari2019enantioseparation}, electro-polymerization processes  \cite{tassinari2020spin,bhowmick2022spin}, and self-assembly processes \cite{al2022spin,stovbun2020weak} were significantly different on differently magnetized substrates.  In most cases, an amplification mechanism is needed to achieve effects over large timescales.

In the context of spin-controlled self-assembly, the aggregation of proteins into amyloid-like fibrils is particularly noteworthy for three reasons. First, intercepting nucleation events is critical for addressing amyloid-related diseases, and while functionalized surfaces have been employed for this purpose,\cite{grigolato2021role} magnetic substrates remain unexplored. Second, amyloids exhibit pronounced stereoselectivity,\cite{brack1980beta} which may serve as an amplification mechanism for spin-dependent aggregation. Third, the A${\beta}$ polypeptide carries a significant charge, and its fibrils form a strong electrical dipole,\cite{muscat2020elucidating} further highlighting their potential in spin-selective processes.

In this study, we demonstrate how spin interactions influence protein aggregation into amyloid fibrils. By leveraging magnetic substrates, we establish that the amount of fibril formation is strongly influenced by substrate magnetization direction and monomer chirality, its size, and electrical dipole or polarization. Additional structural analysis reveals structural changes in the resulting fibrils due to the substrate magnetization. These findings highlight the importance of transient spin polarization, mediated by the CISS effect, in self-assembly processes.

\subsection*{Amyloid ${\beta}$ (1-42) Aggregation on Magnetic Substrates}

\begin{figure}[H]
    \centering \includegraphics{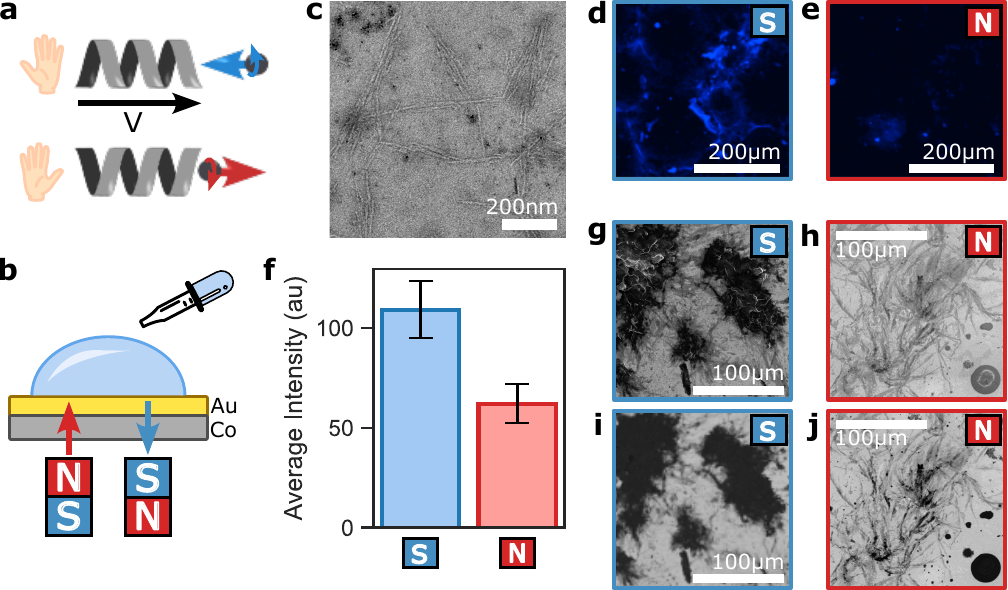}
    \caption[Amyloid ${\beta}$ (1-42)]{\textbf{a} splacement depends on their spin due to the Chiral-Induced Spin Selectivity (CISS) effect, leading to transient spin polarization in the molecule. \textbf{b} Experimental setup: A 40 $\mu$M solution of A$\beta$ polypeptide in PBS was drop-casted onto an Al$_2$O$_3$(0001)/Pt(5 nm)/Au(20 nm)/Co(1.3 nm)/Au(50 nm) substrate, magnetized either North (red) or South (blue) in perpendicular direction to the surface, incubated, and dried at 37°C to allow amyloid-like fibril formation. \textbf{c} TEM image showing negatively stained fibrillar aggregates approximately 200 nm in length and 10 nm in diameter. d-e Fluorescence microscopy images of amyloid-like fibrils stained with ThT, excited at 405 nm and collected at 450-550 nm. South magnetization direction (\textbf{d}) exhibits increased amyloid formation compared to North magnetization direction (\textbf{e}), as indicated by the higher fluorescence intensity. \textbf{f} Quantification of average fluorescence intensity shows significantly higher counts values for South magnetization (157 $\pm$ 14) compared to North direction (110 $\pm$ 10). Errors represent the standard error of the mean. \textbf{g-j} SEM secondary electron images (\textbf{g, h}) and corresponding backscattered electron images (\textbf{i, j}) reveal denser aggregation for South magnetization direction(\textbf{g, i)}), consistent with fluorescence results. Backscattered images (\textbf{i, j}) differentiate between the surface (Au – bright area), protein (dark), and residual salt from the solution (mid-gray in \textbf{j}).}
    \label{Abeta}
\end{figure}

The effect of substrate magnetization direction on the amyloid-like aggregation of Amyloid $\beta$ (1-42) (A$\beta$) polypeptide was studied using the experimental setup illustrated in \autoref{Abeta}b. A solution of A$\beta$ 1-42 protein in Phosphate-buffered saline (PBS) was drop-casted onto Al$_2$O$_3$(0001)/Pt(5 nm)/Au(20 nm)/Co(1.3 nm)/Au(50 nm) MBE grown epitaxial nanostructure used as a substrate. The above nanostructures possess a perpendicular magnetic anisotropy (PMA), making them well-suited for magnetization reorientation by a short pulse of low magnetic field applied out of the plane, therefore avoiding induced undesired magnetic field effects on the assembly process. The substrates were magnetized in either a North (red) or South (blue) direction perpendicular to the plane using a permanent magnet. The comparison between the two magnetization directions was conducted using a combination of confocal fluorescence microscopy with Thioflavin T (ThT) dye, Scanning Electron Microscopy (SEM), and Transmission Electron Microscopy (TEM) to examine their quantity and morphology. The samples were dried at 37°C on the magnetic substrate to allow amyloid fibril formation. Concurrently, the leftover protein solution was incubated at 37°C for 24 hours and dried at a Transmission Electron Microscopy (TEM) grid for the structural analysis. 

The TEM imaging of the negatively stained sample revealed that the A$\beta$ polypeptide self-assembled into fibrillar structures approximately 200 nm in length and 10 nm in diameter (\autoref{Abeta}c). Further analysis of the structure using Attenuated Total Reflection Fourier Transform Infrared (FTIR) revealed that the fibrillar structures are indeed Amyloid-like and composed of $\beta$ sheets, as discussed further in the structural analysis section. Confocal fluorescence microscopy using ThT dye, excited at 405 nm and collected at 480 nm, was employed to detect and quantify the number of amyloid-like fibrils on substrates magnetized in either the North or South directions, respectively. ThT is a fluorescent dye commonly used in microscopy, it specifically binds to amyloid fibrils and enhances its fluorescence intensity. The fluorescence intensity is thus proportional to the total of amyloid-like structures in the studied sample (see Supporting Information for more details). 

To ensure robust sampling, two independent samples were prepared for each magnetization direction, and three 635 $\mu$m $\times$ 635 $\mu$m regions per sample were analyzed, avoiding edges to minimize the effect of the coffee ring. The fluorescence intensity of each image was then taken and averaged between the different samples and regions. The baseline ThT fluorescence in PBS (48) was reduced from the average intensity. Given the 3 mm $\times$ 3 mm size of the drop-cast area, this approach allows imaging of most of the deposited sample. Obviously, the solution concentration is the same for all measured samples. Quantification of average fluorescence intensity (\autoref{Abeta}f) revealed values of 109$\pm$14 counts for South and 62$\pm$10 counts for North magnetization orientation. This nearly twofold increase in intensity indicates higher amyloid aggregation for South-magnetized substrates and strongly supports that substrate magnetization influences the protein self-assembly process. 

SEM secondary electron images (\autoref{Abeta}g, h) and corresponding SEM backscattered electron analysis (\autoref{Abeta}i, j) provide insight into the morphology of amyloid aggregates formed on magnetic substrates. The SEM images reveal that the short fibrils observed in TEM (\autoref{Abeta}c) assemble into larger clumps on the substrate. However, the SEM technique does not resolve the fine details of individual fibrils visible in TEM, focusing instead on the larger- aggregation scale.

In backscattered electron images, intensity is directly related to the atomic number of the material, while higher atomic number elements appear brighter due to stronger electron scattering, but elements with lower atomic number appear darker. In the backscattered images (\autoref{Abeta}i, j), the bright regions correspond to the Au cover layer of the substrate, the dark regions to A$\beta$ polypeptide aggregates, and the mid-gray regions to residual salt from the solution. This analysis reveals that much of the material visible in SEM images for the North magnetization direction (\autoref{Abeta}h) is salt, indicating that the amyloid aggregation itself is less dense than might initially appear. Consistent with the fluorescence microscopy results, the South-magnetized samples exhibit denser amyloid-like aggregation (\autoref{Abeta}g, i) compared to North-magnetized samples (\autoref{Abeta}h, j).

\subsection*{Breaking Down the Protein - the Role of the Monomer's Dipole Moment}

\begin{figure}[H]
    \centering \includegraphics{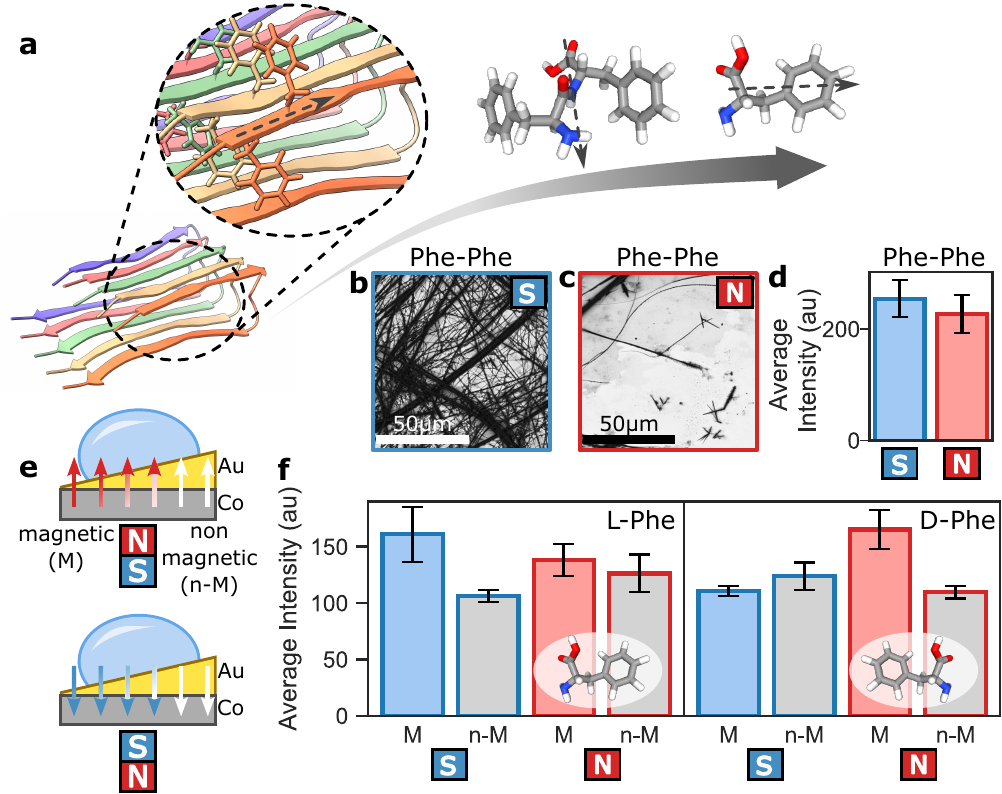}
    \caption[smaller monomers]{\textbf{a} Schematic representation of the A$\beta$ protein and its breakdown into Phe-Phe peptide and single Phe amino acid. The direction of the dipole moment is marked with a dashed arrow. \textbf{b} SEM image of Phe-Phe fibrils formed on a South- (\textbf{b}) or North (\textbf{c}) magnetized substrate. Significantly longer fibrils formed on the South magnetized sample. \textbf{d} Corresponding average fluorescence intensity. \textbf{e} Assembly of L-Phe or D-Phe amino acids drop-casted onto a magnetic substrate magnetized in North (red) or South (blue) direction. Magnetic (M) and nonmagnetic (n-M) regions are created by varying the Au layer thickness on a Co substrate (Au cover layer was deposited as a wedge), allowing comparison of magnetic effects within the same drop. \textbf{f} Average fluorescence intensity for L-Phe shows higher values for the South-magnetized sample (M S = 161 $\pm$ 24) compared to the North-magnetized sample (M N = 139 $\pm$ 15) and nonmagnetic sides (n-M S = 107 $\pm$ 5, n-M N = 127 $\pm$ 17). For D-Phe, the North-magnetized sample (M N = 166 $\pm$ 17) exhibits higher intensity than the South-magnetized sample (M S = 111 $\pm$ 4), with comparable values for the nonmagnetic areas (n-M N = 110 $\pm$ 6, n-M S = 124 $\pm$ 12). Error bars represent standard errors of the mean.}
    \label{Breaking}
\end{figure}

We explored the role of monomer dipole moments in spin-dependent self-assembly. The A$\beta$ 1-42 polypeptide is fragmented into two smaller monomers, Phe-Phe dipeptide, and single amino acid, Phe, as illustrated in \autoref{Breaking}a. The structure is taken from Luhrs et al. \cite{luhrs20053d} Breaking down the protein reduces and shifts its overall dipole moment. Phe-Phe, a core A$\beta$ recognition motif, is symmetric and its dipole moment is aligned along the peptide backbone in the direction of fibril growth, unlike Phe, whose dipole is perpendicular, similar to the A$\beta$ polypeptide (gray arrows, \autoref{Breaking}a).

Similarly to the A$\beta$ polypeptide, Phe-Phe (0.1 mg/mL) was drop-casted onto magnetized substrates (Al$_2$O$_3$(0001)/Pt/Au/Co/Au) and left to self-assemble. SEM and ThT fluorescence microscopy quantified fibril morphology and aggregation. When formed on a South-magnetized substrate (\autoref{Breaking}b), the Phe-Phe monomers formed long fibers extending beyond the area of view of the SEM imaging (greater than 250 $\mu$m). In contrast, when formed on a North-magnetized substrate (\autoref{Breaking}c), the Phe-Phe monomers reveal a heterogeneous assembly consisting of long fibers (greater than 250 $\mu$m) and significantly shorter fiber structures (approximately 10 $\mu$m). Using a confocal microscope, the fluorescence intensity of three images across the sample was taken and averaged (\autoref{Breaking}d). The baseline ThT fluorescence in water (82) was subtracted from the average intensity. 

Because of the different dipole moment directions in the Phe-Phe peptide compared to the A$\beta$ polypeptide, an even smaller monomer, a single Phe amino acid, was studied. As illustrated in \autoref{Breaking}a, the dipole moment in the Phe amino acid is pointing toward its aromatic ring, perpendicular to the fibril growth (similar to the protein). The strength of the dipole moment is much smaller than that of the full protein. In addition, to reduce sample-to-sample variations, the effect of different magnetization orientations within a single droplet was studied. 

To achieve this, Amino acid solutions (L-Phe and D-Phe) were drop-cast onto a magnetic substrate with a wedge-shaped Au layer (0–10 nm thick). The Au layer limits spin penetration from the underlying ferromagnetic Co (\autoref{Breaking}e), creating a spin-polarized (M) region on the thin Au side ($<$5 nm) and a non-spin-polarized (n-M) region on the thick Au side ($>$5 nm).\cite{ghosh2020effect}

ThT fluorescence was used to evaluate the number of amyloid-like fibrils at both areas (magnetic (M) and nonmagnetic (n-M)) for each magnetization direction (\autoref{Breaking}f). The left-handed amino acid displays higher fluorescence intensity on the South-magnetized substrate (M S = 161 $\pm$ 24) compared to the same substrate's nonmagnetic area (n-M S = 107 $\pm$ 5). For assemblies on the North-magnetized substrate, comparable fluorescence was observed for the magnetic and nonmagnetic areas (M N = 139 $\pm$ 15, n-M N = 127 $\pm$ 17). Results for the right-handed amino acid yield the opposite effect: higher fluorescence for assembly on the North-magnetized substrate area (M N = 166 $\pm$ 17) compared to the same substrate's nonmagnetic area (n-M N = 110 $\pm$ 6). On the South-magnetized substrate, there was no significant difference between the magnetic and nonmagnetic areas (M S = 111 $\pm$ 4, n-M S = 124 $\pm$ 12).

Overall, the South-magnetized substrate encourages the assembly of the left-handed amino acid, while the North-magnetized substrate encourages the assembly of the right-handed amino acid, displaying a reversal of the spin effect due to the reversal of the chirality. While the two enantiomers have the same electric dipole, their spin polarization is opposite. The errors in fluorescence values are due to the fact that all the data was taken from the same drop, thus influencing the statistics. Moreover, in comparison to the nonmagnetic substrate area, the magnetic substrate area promotes the self-assembly process rather than hindering it.

\subsection*{Structural Analysis of the Fibrils}

\begin{figure}[H]
    \centering \includegraphics{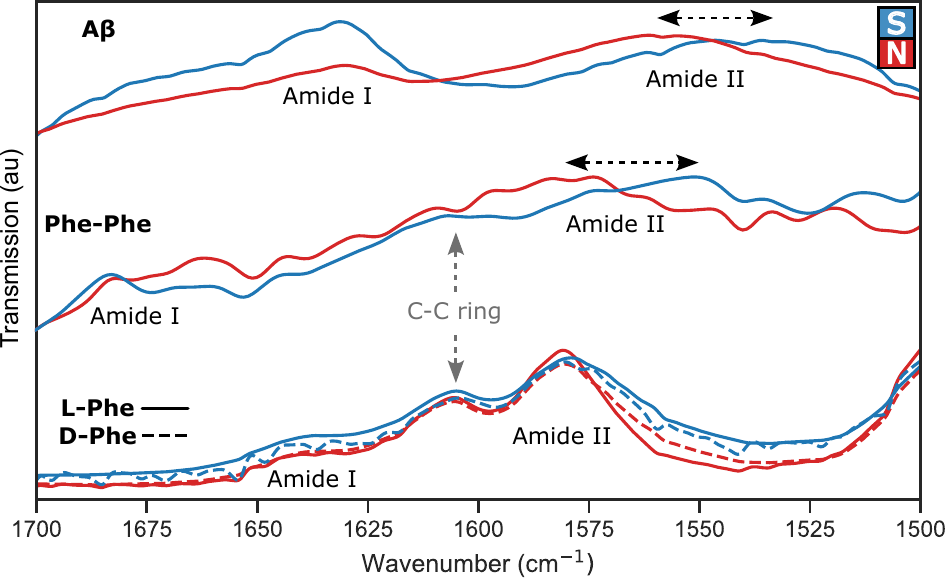}
    \caption[FTIR]{ATR-FTIR spectra of A$\beta$ polypeptide (top), Phe-Phe dipeptide (middle), and L-Phe (solid line) and D-Phe (dashed line) amino acids (bottom) formed on magnetic surfaces magnetized either in North (red) or South (blue) direction. The spectra show characteristic Amide I (1640-1630 cm$^{-1}$) and Amide II (1600-1500 cm$^{-1}$) peaks, indicating a cross-$\beta$ structure, characteristic of amyloid-like fibrils. The Phe and Phe-Phe spectra also show an additional peak at 1605 cm$^{-1}$, indicative of the C-C ring. In A$\beta$, a decrease in the Amide I peak and a 30 wavenumber shift toward higher frequencies are observed for the South to North magnetization directions. In Phe-Phe, only the shift is observed, with no decrease in transmission. In Phe, the spectra remain unchanged, indicating no structural differences between North and South magnetization directions.}
    \label{FTIR}
\end{figure}

ATR-FTIR experiments confirmed the fibril structures. The analysis was focused on Amide I (1700-1600 cm$^{-1}$) and Amide II (1600-1500 cm$^{-1}$) regions. Different spectra were observed due to the difference in the monomers used. The recurring peaks at 1630 cm$^{-1}$ (A$\beta$), 1685 cm$^{-1}$ (Phe-Phe), and 1640 cm$^{-1}$ (Phe) (see \autoref{FTIR}) fit well the known signatures for $\beta$-sheets. \cite{reches2005self,sarroukh2013atr} The Phe and Phe-Phe spectra also show an additional peak at 1605 cm$^{-1}$, indicative of the C-C ring. \cite{barth2000infrared} The Amide II range shows varying peaks (1580-1530 cm$^{-1}$), corroborating cross-$\beta$ structure. 

When comparing the spectra of A$\beta$ prepared on North or South magnetization orientation, two main differences are observed: a decrease in the intensity of the Amide I peak compared to the Amide II peak and a significant 30 wavenumber shift in Amide II peaks. The spectral changes slowly disappear in smaller monomers, Phe-Phe and Phe. In the Phe-Phe spectra, only the Amide II shift is observed, with no decrease. The Phe spectra remain unchanged, indicating no structural differences between North and South magnetization directions.

\subsection*{Discussion}
The changes in amyloid-like self-assembly observed here can be divided into two types, morphological and structural. Fibril morphology varied with substrate magnetization orientation. South magnetization consistently favored fibril formation (more fibrils and longer fibrils) even for different L-chiral monomers. This can be interpreted as changes to the dynamics of the self-assembly process. As the chiral molecule approaches the substrate surface, it undergoes charge reorganization and therefore spin polarization due to the CISS effect. Thus, the spin-polarized substrate and molecule have either parallel (triplet-like) or antiparallel (singlet-like) spins where the antiparallel case is energetically favored. 

Different nucleation rates, as seen in various studies, \cite{tassinari2019enantioseparation,banerjee2018separation,safari2024enantioselective} are common for chiral molecules adsorbed on perpendicularly magnetized substrates and can explain the change in the amount of fibril formation found in this study. However, as spin effects are transient \cite{sukenik2020correlation}, an amplification mechanism is required to sustain the effects over larger timescales. In the case of Tassinari et al., \cite{tassinari2019enantioseparation} this amplification mechanism was a conglomerate crystallization. Interestingly, in our case, a possible amplification mechanism is the stereoselectivity of the beta-sheet. A possible path to test this is to create amyloid fibrils from a racemic mixture and to look for homochiral fibrils depending on the substrate magnetization direction. Moreover, while alpha-helix oligopeptides are a go-to model for CISS due to the helix's clear chiral structure and its connection to the spin polarization degree \cite{gohler2011spin,eckshtain2016cold}, the beta-sheet structure and its spin selective properties remain less explored.

ATR-FTIR spectra revealed a decreased Amide I/II peak ratio and a significant shift in the Amide II peak. These spectral shifts indicate structural differences between oligomers and mature fibrils. \cite{sarroukh2011transformation,sarroukh2013atr} Both species are likely present in the sample, with their relative proportions varying. Notably, despite the fact that South magnetization orientation exhibiting higher fluorescence—suggesting increased amyloid content—its ATR-FTIR signature indicates a higher proportion of oligomers relative to mature fibrils. These findings suggest that spin interactions may play a significant role in protein structural transitions.

While studying the different monomers, several key findings were observed. First, switching chirality alters the spin polarization while the dipole remains unchanged, highlighting the role of chirality in spin interactions. Second, distinct effects were observed on the same surface at different areas (magnetic or nonmagnetic), depending on the presence or absence of spin injection, suggesting a localized influence of spin-polarized currents on the system. Lastly, the role of the monomer's dipole moment was explored, showing that different dipole directions yielded different morphology on the magnetized substrate. Specifically, the Phe-Phe case is interesting, where the monomer dipole moment is in the growth direction. In this case, a significant shortening of the fibrils was observed for the unfavored magnetization direction. 

Spin is inherently challenging to define and measure in biological systems since there is no clear axis, which has historically limited its investigation. However, by conducting the self-assembly process on a magnetic substrate and connecting the spin to the chiral structure of the fibril, it becomes possible to define a specific spin direction. Moreover, spin exchange interactions, while very short-ranged, are extremely strong. Here spin exchange interactions are introduced to the system by a spin polarized thin ferromagnetic layer and the chirality of the monomers resulting in significant effects using low magnetic fields. 

Functionalized substrates offer a unique platform for controlling amyloid self-assembly, as aggregation is highly sensitive to surface properties.\cite{grigolato2021role} As delivery to target sites remains a major challenge, an alternative approach is integrating such surfaces into blood dialysis to treat dialysis-related amyloidosis. Further research is needed regarding the generality of the effect. Its strength will likely depend on the molecule’s electric dipole.

It is important to note that electronic spin effects are rather unexpected in biological systems which are large and scattered. The average length of a single fibril is 200 nm, and the aggregation is 1–100 $\mu$m in size. Both are significantly longer than the expected length for spin effects at these conditions, which is usually in the few \AA  range. \cite{metzger2020electron} This is likely because, in a self-assembly process, dipoles stack on top of each other, enhancing the effectiveness of the CISS effect in the dynamic process.

\subsection*{Conclusion} 
In spite of its key role in amyloid formation, the role of spin polarization in this process was never explored. Our study reveals that transient spin polarization, driven by the Chiral Induced Spin Selectivity (CISS) effect, play a significant role in amyloid-like fibril formation. Using magnetized substrates with perpendicular anisotropy, we observe a preferential magnetization direction for fibril formation, resulting in up to twice as many fibrils formed and significantly longer fibrils in the favored magnetization direction. The monomer chirality and dipole moment are important in determining the favored magnetization orientation. Shifts between oligomers to mature fibrils between samples further suggest that spin interactions influence protein structure. Despite the expected short-range nature of spin interactions, the effects persist at much larger scales. These findings not only provide insights into amyloid formation but also open new possibilities for using magnetic particles in therapeutic strategies for amyloid-related disorders.

\printbibliography

@article{knowles2014amyloid,
  title={The amyloid state and its association with protein misfolding diseases},
  author={Knowles, Tuomas PJ and Vendruscolo, Michele and Dobson, Christopher M},
  journal={Nature reviews Molecular cell biology},
  volume={15},
  number={6},
  pages={384--396},
  year={2014},
  publisher={Nature Publishing Group UK London}
}

@article{wei2017self,
  title={Self-assembling peptide and protein amyloids: from structure to tailored function in nanotechnology},
  author={Wei, Gang and Su, Zhiqiang and Reynolds, Nicholas P and Arosio, Paolo and Hamley, Ian W and Gazit, Ehud and Mezzenga, Raffaele},
  journal={Chemical Society Reviews},
  volume={46},
  number={15},
  pages={4661--4708},
  year={2017},
  publisher={Royal Society of Chemistry}
}

@article{reches2003casting,
  title={Casting metal nanowires within discrete self-assembled peptide nanotubes},
  author={Reches, Meital and Gazit, Ehud},
  journal={Science},
  volume={300},
  number={5619},
  pages={625--627},
  year={2003},
  publisher={American Association for the Advancement of Science}
}

@article{reches2005self,
  title={Self-assembly of peptide nanotubes and amyloid-like structures by charged-termini-capped diphenylalanine peptide analogues},
  author={Reches, Meital and Gazit, Ehud},
  journal={Israel journal of chemistry},
  volume={45},
  number={3},
  pages={363--371},
  year={2005},
  publisher={Wiley Online Library}
}

@article{adler2012phenylalanine,
  title={Phenylalanine assembly into toxic fibrils suggests amyloid etiology in phenylketonuria},
  author={Adler-Abramovich, Lihi and Vaks, Lilach and Carny, Ohad and Trudler, Dorit and Magno, Andrea and Caflisch, Amedeo and Frenkel, Dan and Gazit, Ehud},
  journal={Nature chemical biology},
  volume={8},
  number={8},
  pages={701--706},
  year={2012},
  publisher={Nature Publishing Group US New York}
}

@article{wang2011charged,
  title={Charged diphenylalanine nanotubes and controlled hierarchical self-assembly},
  author={Wang, Minjie and Du, Lingjie and Wu, Xinglong and Xiong, Shijie and Chu, Paul K},
  journal={Acs Nano},
  volume={5},
  number={6},
  pages={4448--4454},
  year={2011},
  publisher={ACS Publications}
}

@article{a2007alignment,
  title={Alignment of aromatic peptide tubes in strong magnetic fields},
  author={A. Hill, RJ and Sedman, Victoria L and Allen, Stephanie and Williams, P and Paoli, Massimo and Adler-Abramovich, Lihi and Gazit, Ehud and Eaves, Laurence and Tendler, Saul JB},
  journal={Advanced Materials},
  volume={19},
  number={24},
  pages={4474--4479},
  year={2007},
  publisher={Wiley Online Library}
}

@article{pandey2017modulation,
  title={Modulation of peptide based nano-assemblies with electric and magnetic fields},
  author={Pandey, Gaurav and Saikia, Jahnu and Sasidharan, Sajitha and Joshi, Deep C and Thota, Subhash and Nemade, Harshal B and Chaudhary, Nitin and Ramakrishnan, Vibin},
  journal={Scientific reports},
  volume={7},
  number={1},
  pages={2726},
  year={2017},
  publisher={Nature Publishing Group UK London}
}

@article{zaguri2021kinetic,
  title={Kinetic and thermodynamic driving factors in the assembly of phenylalanine-based modules},
  author={Zaguri, Dor and Zimmermann, Manuela R and Meisl, Georg and Levin, Aviad and Rencus-Lazar, Sigal and Knowles, Tuomas PJ and Gazit, Ehud},
  journal={ACS nano},
  volume={15},
  number={11},
  pages={18305--18311},
  year={2021},
  publisher={ACS Publications}
}

@article{su2023electric,
  title={Electric field-assisted self-assembly of diphenylalanine peptides for high-performance energy conversion},
  author={Su, Yusen and Liu, Jing and Yang, Dingyi and Hu, Wen and Jiang, Xue and Wang, Zhong Lin and Yang, Rusen},
  journal={ACS Materials Letters},
  volume={5},
  number={9},
  pages={2317--2323},
  year={2023},
  publisher={ACS Publications}
}

@article{nguyen2016self,
  title={Self-assembly of diphenylalanine peptide with controlled polarization for power generation},
  author={Nguyen, Vu and Zhu, Ren and Jenkins, Kory and Yang, Rusen},
  journal={Nature communications},
  volume={7},
  number={1},
  pages={13566},
  year={2016},
  publisher={Nature Publishing Group UK London}
}

@article{gohler2011spin,
  title={Spin selectivity in electron transmission through self-assembled monolayers of double-stranded DNA},
  author={G{\"o}hler, B and Hamelbeck, V and Markus, TZ and Kettner, M and Hanne, GF and Vager, Zeev and Naaman, Ron and Zacharias, H},
  journal={Science},
  volume={331},
  number={6019},
  pages={894--897},
  year={2011},
  publisher={American Association for the Advancement of Science}
}

@article{xie2011spin,
  title={Spin specific electron conduction through DNA oligomers},
  author={Xie, Zouti and Markus, Tal Z and Cohen, Sidney R and Vager, Zeev and Gutierrez, Rafael and Naaman, Ron},
  journal={Nano letters},
  volume={11},
  number={11},
  pages={4652--4655},
  year={2011},
  publisher={ACS Publications}
}

@article{mishra2013spin,
  title={Spin-dependent electron transmission through bacteriorhodopsin embedded in purple membrane},
  author={Mishra, Debabrata and Markus, Tal Z and Naaman, Ron and Kettner, Matthias and G{\"o}hler, Benjamin and Zacharias, Helmut and Friedman, Noga and Sheves, Mordechai and Fontanesi, Claudio},
  journal={Proceedings of the National Academy of Sciences},
  volume={110},
  number={37},
  pages={14872--14876},
  year={2013},
  publisher={National Academy of Sciences}
}

@article{niman2023bacterial,
  title={Bacterial extracellular electron transfer components are spin selective},
  author={Niman, Christina M and Sukenik, Nir and Dang, Tram and Nwachukwu, Justus and Thirumurthy, Miyuki A and Jones, Anne K and Naaman, Ron and Santra, Kakali and Das, Tapan K and Paltiel, Yossi and others},
  journal={The Journal of chemical physics},
  volume={159},
  number={14},
  year={2023},
  publisher={AIP Publishing}
}

@article{varade2018bacteriorhodopsin,
  title={Bacteriorhodopsin based non-magnetic spin filters for biomolecular spintronics},
  author={Varade, Vaibhav and Markus, Tal and Vankayala, Kiran and Friedman, Noga and Sheves, Mordechai and Waldeck, David H and Naaman, Ron},
  journal={Physical Chemistry Chemical Physics},
  volume={20},
  number={2},
  pages={1091--1097},
  year={2018},
  publisher={Royal Society of Chemistry}
}

@article{gupta2023spin,
  title={Spin-dependent electrified protein interfaces for probing the CISS effect},
  author={Gupta, Ritu and Chinnasamy, Hariharan V and Sahu, Dipak and Matheshwaran, Saravanan and Sow, Chanchal and Chandra Mondal, Prakash},
  journal={The Journal of Chemical Physics},
  volume={159},
  number={2},
  year={2023},
  publisher={AIP Publishing}
}

@article{sang2021temperature,
  title={Temperature dependence of charge and spin transfer in azurin},
  author={Sang, Yutao and Mishra, Suryakant and Tassinari, Francesco and Karuppannan, Senthil Kumar and Carmieli, Raanan and Teo, Ruijie D and Migliore, Agostino and Beratan, David N and Gray, Harry B and Pecht, Israel and others},
  journal={The Journal of Physical Chemistry C},
  volume={125},
  number={18},
  pages={9875--9883},
  year={2021},
  publisher={ACS Publications}
}

@article{aminadav2024chiral,
  title={Chiral Nematic Cellulose Nanocrystal Films for Enhanced Charge Separation and Quantum-Confined Stark Effect},
  author={Aminadav, Gur and Shoseyov, Omer and Belsey, Shylee and Voignac, Daniel and Yochelis, Shira and Levi-Kalisman, Yael and Yan, Binghai and Shoseyov, Oded and Paltiel, Yossi},
  journal={ACS nano},
  volume={18},
  number={42},
  pages={28609--28621},
  year={2024},
  publisher={ACS Publications}
}

@article{tassinari2019enantioseparation,
  title={Enantioseparation by crystallization using magnetic substrates},
  author={Tassinari, Francesco and Steidel, Jakob and Paltiel, Shahar and Fontanesi, Claudio and Lahav, Meir and Paltiel, Yossi and Naaman, Ron},
  journal={Chemical science},
  volume={10},
  number={20},
  pages={5246--5250},
  year={2019},
  publisher={Royal Society of Chemistry}
}

@article{lu2021enantiospecificity,
  title={Enantiospecificity of cysteine adsorption on a ferromagnetic surface: Is it kinetically or thermodynamically controlled?},
  author={Lu, Y and Bloom, BP and Qian, S and Waldeck, DH},
  journal={The Journal of Physical Chemistry Letters},
  volume={12},
  number={32},
  pages={7854--7858},
  year={2021},
  publisher={ACS Publications}
}

@article{safari2024enantioselective,
  title={Enantioselective adsorption on magnetic surfaces},
  author={Safari, Mohammad Reza and Matthes, Frank and Caciuc, Vasile and Atodiresei, Nicolae and Schneider, Claus M and Ernst, Karl-Heinz and B{\"u}rgler, Daniel E},
  journal={Advanced Materials},
  volume={36},
  number={14},
  pages={2308666},
  year={2024},
  publisher={Wiley Online Library}
}

@article{kumar2017chirality,
  title={Chirality-induced spin polarization places symmetry constraints on biomolecular interactions},
  author={Kumar, Anup and Capua, Eyal and Kesharwani, Manoj K and Martin, Jan ML and Sitbon, Einat and Waldeck, David H and Naaman, Ron},
  journal={Proceedings of the National Academy of Sciences},
  volume={114},
  number={10},
  pages={2474--2478},
  year={2017},
  publisher={National Academy of Sciences}
}

@article{wei2023examining,
  title={Examining the effects of homochirality for electron transfer in protein assemblies},
  author={Wei, Jimeng and Bloom, Brian P and Dunlap-Shohl, Wiley A and Clever, Caleb B and Rivas, Jos{\'e} E and Waldeck, David H},
  journal={The Journal of Physical Chemistry B},
  volume={127},
  number={29},
  pages={6462--6469},
  year={2023},
  publisher={ACS Publications}
}

@article{tassinari2020spin,
  title={Spin-dependent enantioselective electropolymerization},
  author={Tassinari, Francesco and Amsallem, Dana and Bloom, Brian P and Lu, Yiyang and Bedi, Anjan and Waldeck, David H and Gidron, Ori and Naaman, Ron},
  journal={The Journal of Physical Chemistry C},
  volume={124},
  number={38},
  pages={20974--20980},
  year={2020},
  publisher={ACS Publications}
}

@article{bhowmick2022spin,
  title={Spin-induced asymmetry reaction—The formation of asymmetric carbon by electropolymerization},
  author={Bhowmick, Deb Kumar and Das, Tapan Kumar and Santra, Kakali and Mondal, Amit Kumar and Tassinari, Francesco and Schwarz, Rony and Diesendruck, Charles E and Naaman, Ron},
  journal={Science advances},
  volume={8},
  number={31},
  pages={eabq2727},
  year={2022},
  publisher={American Association for the Advancement of Science}
}

@article{grigolato2021role,
  title={The role of surfaces on amyloid formation},
  author={Grigolato, Fulvio and Arosio, Paolo},
  journal={Biophysical Chemistry},
  volume={270},
  pages={106533},
  year={2021},
  publisher={Elsevier}
}

@article{brack1980beta,
  title={$\beta$-structures of polypeptides with L-and D-residues: Part III. Experimental Evidences for Enrichment in Enantiomer},
  author={Brack, Andr{\'e} and Spach, G{\'e}rard},
  journal={Journal of Molecular Evolution},
  volume={15},
  pages={231--238},
  year={1980},
  publisher={Springer}
}

@article{muscat2020elucidating,
  title={Elucidating the effect of static electric field on amyloid beta 1--42 supramolecular assembly},
  author={Muscat, S and Stojceski, F and Danani, A},
  journal={Journal of Molecular Graphics and Modelling},
  volume={96},
  pages={107535},
  year={2020},
  publisher={Elsevier}
}

@article{luhrs20053d,
  title={3D structure of Alzheimer's amyloid-$\beta$ (1--42) fibrils},
  author={L{\"u}hrs, Thorsten and Ritter, Christiane and Adrian, Marc and Riek-Loher, Dominique and Bohrmann, Bernd and D{\"o}beli, Heinz and Schubert, David and Riek, Roland},
  journal={Proceedings of the National Academy of Sciences},
  volume={102},
  number={48},
  pages={17342--17347},
  year={2005},
  publisher={National Academy of Sciences}
}

@article{ghosh2020effect,
  title={Effect of chiral molecules on the electron’s spin wavefunction at interfaces},
  author={Ghosh, Supriya and Mishra, Suryakant and Avigad, Eytan and Bloom, Brian P and Baczewski, LT and Yochelis, Shira and Paltiel, Yossi and Naaman, Ron and Waldeck, David H},
  journal={The journal of physical chemistry letters},
  volume={11},
  number={4},
  pages={1550--1557},
  year={2020},
  publisher={ACS Publications}
}

@article{sukenik2020correlation,
  title={Correlation between ferromagnetic layer easy axis and the tilt angle of self assembled chiral molecules},
  author={Sukenik, Nir and Tassinari, Francesco and Yochelis, Shira and Millo, Oded and Baczewski, Lech Tomasz and Paltiel, Yossi},
  journal={Molecules},
  volume={25},
  number={24},
  pages={6036},
  year={2020},
  publisher={MDPI}
}

@article{eckshtain2016cold,
  title={Cold denaturation induces inversion of dipole and spin transfer in chiral peptide monolayers},
  author={Eckshtain-Levi, Meital and Capua, Eyal and Refaely-Abramson, Sivan and Sarkar, Soumyajit and Gavrilov, Yulian and Mathew, Shinto P and Paltiel, Yossi and Levy, Yaakov and Kronik, Leeor and Naaman, Ron},
  journal={Nature communications},
  volume={7},
  number={1},
  pages={10744},
  year={2016},
  publisher={Nature Publishing Group UK London}
}

@article{sarroukh2013atr,
  title={ATR-FTIR: A “rejuvenated” tool to investigate amyloid proteins},
  author={Sarroukh, Rabia and Goormaghtigh, Erik and Ruysschaert, Jean-Marie and Raussens, Vincent},
  journal={Biochimica et Biophysica Acta (BBA)-Biomembranes},
  volume={1828},
  number={10},
  pages={2328--2338},
  year={2013},
  publisher={Elsevier}
}

@article{tjernberg1996arrest,
  title={Arrest of-Amyloid Fibril Formation by a Pentapeptide Ligand (∗)},
  author={Tjernberg, Lars O and N{\"a}slund, Jan and Lindqvist, Fredrik and Johansson, Jan and Karlstr{\"o}m, Anders R and Thyberg, Johan and Terenius, Lars and Nordstedt, Christer},
  journal={Journal of Biological Chemistry},
  volume={271},
  number={15},
  pages={8545--8548},
  year={1996},
  publisher={ASBMB}
}

@article{soto1998beta,
  title={$\beta$-sheet breaker peptides inhibit fibrillogenesis in a rat brain model of amyloidosis: implications for Alzheimer's therapy},
  author={Soto, Claudio and Sigurdsson, Einar M and Morelli, Laura and Asok Kumar, R and Casta{\~n}o, Eduardo M and Frangione, Blas},
  journal={Nature medicine},
  volume={4},
  number={7},
  pages={822--826},
  year={1998},
  publisher={Nature Publishing Group US New York}
}

@article{al2022spin,
  title={Spin-induced organization of cellulose nanocrystals},
  author={Al-Bustami, Hammam and Belsey, Shylee and Metzger, Tzuriel and Voignac, Daniel and Yochelis, Shira and Shoseyov, Oded and Paltiel, Yossi},
  journal={Biomacromolecules},
  volume={23},
  number={5},
  pages={2098--2105},
  year={2022},
  publisher={ACS Publications}
}

@article{naaman2019chiral,
  title={Chiral molecules and the electron spin},
  author={Naaman, Ron and Paltiel, Yossi and Waldeck, David H},
  journal={Nature Reviews Chemistry},
  volume={3},
  number={4},
  pages={250--260},
  year={2019},
  publisher={Nature Publishing Group UK London}
}

@article{banerjee2020long,
  title={Long-range charge reorganization as an allosteric control signal in proteins},
  author={Banerjee-Ghosh, Koyel and Ghosh, Shirsendu and Mazal, Hisham and Riven, Inbal and Haran, Gilad and Naaman, Ron},
  journal={Journal of the American Chemical Society},
  volume={142},
  number={48},
  pages={20456--20462},
  year={2020},
  publisher={ACS Publications}
}

@article{banerjee2018separation,
  title={Separation of enantiomers by their enantiospecific interaction with achiral magnetic substrates},
  author={Banerjee-Ghosh, Koyel and Ben Dor, Oren and Tassinari, Francesco and Capua, Eyal and Yochelis, Shira and Capua, Amir and Yang, See-Hun and Parkin, Stuart SP and Sarkar, Soumyajit and Kronik, Leeor and others},
  journal={Science},
  volume={360},
  number={6395},
  pages={1331--1334},
  year={2018},
  publisher={American Association for the Advancement of Science}
}

@article{kettner2015spin,
  title={Spin filtering in electron transport through chiral oligopeptides},
  author={Kettner, M and Gohler, B and Zacharias, H and Mishra, D and Kiran, V and Naaman, R and Fontanesi, Claudio and Waldeck, David H and Sęk, S{\l}awomir and Paw{\l}owski, Jan and others},
  journal={The Journal of Physical Chemistry C},
  volume={119},
  number={26},
  pages={14542--14547},
  year={2015},
  publisher={ACS Publications}
}

@article{stovbun2020weak,
  title={The weak magnetic field inhibits the supramolecular self-ordering of chiral molecules},
  author={Stovbun, Sergey V and Zanin, Anatoly M and Skoblin, Aleksey A and Zlenko, Dmitry V},
  journal={Scientific Reports},
  volume={10},
  number={1},
  pages={17072},
  year={2020},
  publisher={Nature Publishing Group UK London}
}

@article{metzger2020electron,
  title={The electron spin as a chiral reagent},
  author={Metzger, Tzuriel S and Mishra, Suryakant and Bloom, Brian P and Goren, Naama and Neubauer, Avner and Shmul, Guy and Wei, Jimeng and Yochelis, Shira and Tassinari, Francesco and Fontanesi, Claudio and others},
  journal={Angewandte Chemie},
  volume={132},
  number={4},
  pages={1670--1675},
  year={2020},
  publisher={Wiley Online Library}
}

@article{barth2000infrared,
  title={The infrared absorption of amino acid side chains},
  author={Barth, Andreas},
  journal={Progress in biophysics and molecular biology},
  volume={74},
  number={3-5},
  pages={141--173},
  year={2000},
  publisher={Elsevier}
}

@article{sarroukh2011transformation,
  title={Transformation of amyloid $\beta$ (1--40) oligomers into fibrils is characterized by a major change in secondary structure},
  author={Sarroukh, Rabia and Cerf, Emilie and Derclaye, Sylvie and Dufr{\^e}ne, Yves F and Goormaghtigh, Erik and Ruysschaert, Jean-Marie and Raussens, Vincent},
  journal={Cellular and Molecular Life Sciences},
  volume={68},
  pages={1429--1438},
  year={2011},
  publisher={Springer}
}

\subsection*{Acknowledgments}

\textbf{Funding:}
Y.P. acknowledges funding from the Carl Zeiss Stiftung through the HYMMS project (No. P2022-03-044), the U.S. Air Force Office of Scientific (grant No. FA8655-24-1-7390) and the Israel Science Foundation (grant No. 360/24). E.G. acknowledges support from the Air Force Office of Scientific Research under award No. FA8655-21-1-7004. Y.K. thanks the Israel Council of Higher Education for support through the VTT fellowship for women in STEM.

\textbf{Author contributions:}
Conceptualization: EG, YP, YK
Methodology: YK, GFZ, DM, SY
Investigation: YK, GFZ, DM, OB, NMB, YLK, IT, SY 
Resources: LTB
Supervision: EG, DS, YP
Writing – original draft: YK, DM, GFZ
Writing – review and editing: YK, DM, GFZ, OB, NMB, YLK, IT, SY, LTB, DS, EG, YP

\textbf{Competing interests:} Authors declare that they have no competing interests.

\textbf{Data and materials availability:} All data are available in the main text or the supplementary materials. Full confocal image files are available at  https://doi.org/10.5281/zenodo.14925254

\textbf{Supplementary Materials:}
Materials and Methods, Supplementary Text and Figs. S1 to S11

\newpage
\begin{center}
    \Huge \textbf{Supporting Information}
\end{center}

\tableofcontents
\newpage

\section{Methods}
\subsection{Materials}
The L-phe, D-phe, and diphenylalanine were purchased from Sigma. The diphenylalanine analog Ac-Phe-Phe-NH$_2$ and amyloid \(\beta\)-polypeptide (HFIP-treated) were purchased from Bachem.

\subsection{Sample Preparation}
Allocates of the monomerized A-$\beta$ (1-42) polypeptide were prepared by dissolving 1mg A-$\beta$ (1-42) in 1mL of high-grade 1,1,1,3,3,3-hexafluoroisopropanol (HFIP, Sigma) by 20s sonication on ice then by constant shaking at 150 rpm at 37 C for 90 minutes. The samples were then aliquoted into 200$\mu$g stocks, and the solvent was left to evaporate before storage at -20 C until use.

Fresh stocks of A-$\beta$ (1-42) polypeptide solution were prepared by dissolving 200$\mu$g A-$\beta$ (1-42) in 42.9$\mu$L dimethyl sulfoxide (DMSO, Sigma) by 60s sonication. The solution was diluted in 1066 $\mu$L PBS (pH 7.4) to achieve a final concentration of 40$\mu$M. The solution was dyed with Thioflavin T (ThT, Sigma) at a concentration of 40$\mu$M. 5$\mu$L drops of the final solution were drop-casted onto magnetic substrates, magnetized by placement on a permanent magnet oriented to South or North directions. The droplet was then allowed to dry and self-assemble into fibrils at 37 C.

Fresh stock solutions of the diphenylalanine were prepared by dissolving the lyophilized peptide in HFIP at a concentration of 100 mg/mL. The peptide stock solution was diluted to a final concentration of 2 mg/mL. \cite{reches2005self} The solution was dyed with ThT at a concentration of 10 $\mu$M.

Fresh stock solutions of all other peptides and amino acids were prepared by dissolving the lyophilized peptides in ddH$_2$O at 90 C for 3 hours and diluting to a final concentration of 0.1 mg/mL similar to \cite{adler2012phenylalanine}. The solution was dyed with ThT at a concentration of 10 $\mu$M.

To avoid any pre-aggregation, fresh stock solutions were prepared for each experiment. The final solutions were drop-casted (5$\mu$L drops) onto magnetic substrates, magnetized by placement on a permanent magnet (75 mT) oriented to South or North directions. The droplet was then allowed to dry and self-assemble into fibrils at room temperature.

\subsection{Magnetic Substrates Growth}

Molecular beam epitaxy (MBE) grown epitaxial thin film magnetic samples with perpendicular anisotropy (Al$_2$O$_3$ (0001)/ Pt(50\AA) / Au(200\AA) / Co(13\AA) / Au(50\AA)), were used for the experiments. The FM samples were magnetized by an external magnetic field of 75mT at room temperature. The coercive field of the FM samples used was $\sim$25 mT or $\sim$12 mT (for the gradient samples). The FM samples’ easy axis was out-of-plane (OOP), thus ensuring that the applied magnetic field would reorient the magnetization OOP, parallel or anti-parallel to surface normal.

Epitaxial Au wedge samples of the same cobalt thickness were also grown by MBE with a configuration: Al$_2$O$_3$(0001) / Pt(50\AA) / Au(200\AA) / Co(13\AA) / Au(0-100\AA). The magnetic (M) areas were measured at the 2–5 nm Au thickness area, while the nonmagnetic (n-M) areas were measured at 6–8 nm thickness of the Au layer. The small variations in Au thickness areas studied between different samples were due to slightly different locations of the solution droplet.

For the FTIR on Phe solutions and the experiments with the peptide analogue, Ti (2 nm)/Ni (80 nm)/Au (5nm) film was grown using AJA ATC Polaris Series UHV sputtering system (base Pressure: 3 $\times$ 10$^{-9}$ Torr) onto a Si/SiO wafer.

\subsection{Scanning Electron Microscopy (SEM)}

Scanning Electron Microscopy (SEM) images for the A-$\beta$ polypeptide and Phe-Phe peptide samples were acquired using an Extra-High Resolution Scanning Electron Microscope Magellan 400L (ThermoFisher, formerly FEI). The imaging was conducted at a current of 25 pA and an accelerating voltage of 5 kV, with a working distance of 4.1 mm and a magnification of 350X. Both secondary electron and backscattered electron imaging modes were used. For the Phe amino acid samples, imaging was performed using an Analytical High Resolution Scanning Electron Microscope Apreo 2S (Thermo Fisher Scientific) at a current of 0.1 nA and an accelerating voltage of 2 kV, with a working distance of 4.6 mm and a magnification of 350X.

\subsection{Transmission Electron Microscopy (TEM)}

Samples (10 $\mu$L drop) were placed on a glow discharged carbon-coated 300 mesh copper TEM grids (Ted Pella, Inc.). After blotting, the samples were either dried in air before observation or negatively stained with 2$\%$ aqueous solution of uranyl acetate for 2 minutes and air-dried. The samples were examined by a FEI Tecnai 12 G2 TWIN TEM operated at 120kV. Images were recorded using a 4k $\times$ 4k FEI Eagle CCD camera.

\subsection{Fourier Transform Infrared Spectroscopy}

To assess conformational changes in the fibril structure between the different samples, ATR-FTIR spectra of the amyloid-like structures were recorded using Thermo Scientific Nicolet iS50 FT-IR spectrometer with an ATR accessory with a MCT detector, using a resolution of 4 cm$^{-1}$ and averaging 32 scans per spectrum over the range of 4000–600 cm$^{-1}$.

\subsection{Thioflavin T Binding Assay}

ThT was imaged using the FV-1200 confocal microscope with a 10X/0.45 objective (Olympus, Japan), excited at 405nm and collected at 450-550nm. A bright field image was collected as well. Z-stack images were obtained at 2$\mu$m distance. Time-lapse was also obtained for monitoring the assembly of fibrils.

\section{Sample Images}

\begin{figure}[h]
    \centering
    \includegraphics{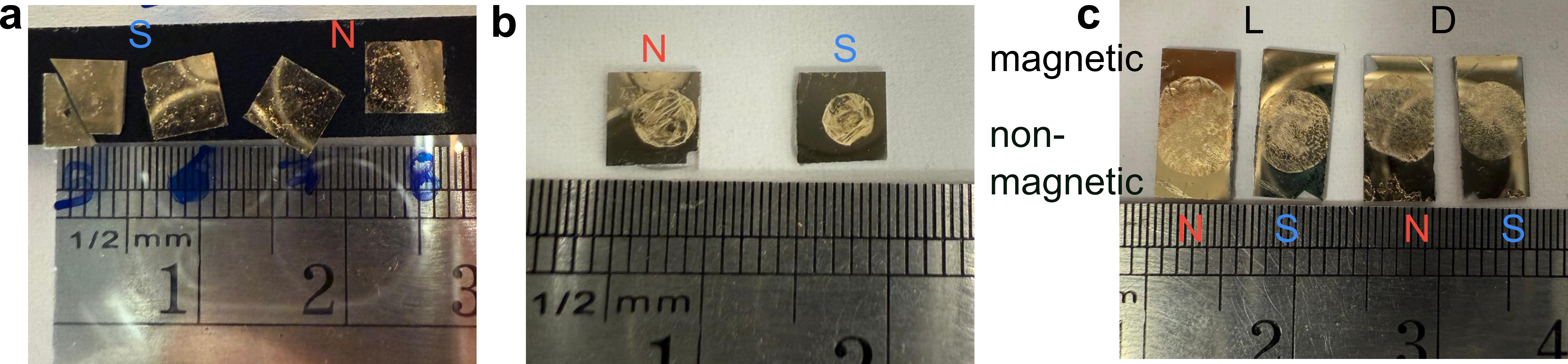}
    \caption[samples]{Optical images of the samples. 
    \textbf{a} A-$\beta$ polypeptide solution dried on an Al$_2$O$_3$ (0001)/Pt(50\AA)/Au(200\AA)/Co(13\AA)/Au(50\AA) substrate magnetized in the North or South direction. 
    \textbf{b} Phe-Phe peptide solution dried on an Al$_2$O$_3$ (0001)/Pt(50\AA)/Au(200\AA)/Co(13\AA)/Au(50\AA) substrate magnetized in the North or South direction. 
    \textbf{c} Phe amino acid solution dried on an Al$_2$O$_3$ (0001)/Pt(50\AA)/Au(200\AA)/Co(13\AA)/Au(0-100\AA) substrate magnetized in the North or South direction. The thick (thin) Au side is marked as non-magnetic (magnetic).}
    \label{samples}
\end{figure}

\section{Characterization of Magnetic Substrates}
\begin{figure}[H] 
\centering 
\includegraphics{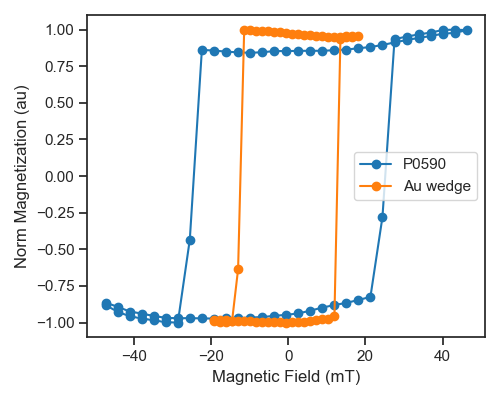} \caption[MOKE sweep]{ Magnetic hysteresis loops of Al$_2$O$_3$ (0001)/Pt(50\AA)/Au(200\AA)/Co(13\AA)/Au(50\AA) (P0590) (blue) and Au wedge (orange) substrate at Au thickness of 4 \AA measured using MOKE microscope} 
\label{MOKE sweep} 
\end{figure}

The magnetic properties of the substrates were characterized using a magneto-optical Kerr effect (MOKE) microscope. MOKE measurements were performed using a commercial Evico Magnetics GmbH magneto-optical Kerr microscope, equipped with an electromagnet and a piezo controller for mechanical stabilization. Magnetic hysteresis loops were obtained by sweeping the applied magnetic field from negative to positive and back to negative. The measured hysteresis loops are shown in \autoref{MOKE sweep}, with the blue curve corresponding to the Al$_2$O$_3$ (0001)/Pt(50\AA)/Au(200\AA)/Co(13\AA)/Au(50\AA) (P0590) substrate and the orange curve representing the Au wedge substrate measured for Au thickness of 4 \AA. The P0590 substrate exhibited a coercive field of 25 mT, while the wedge substrate had a lower coercive field of 12 mT.

\section{Spin Selectivity}
To quantify spin selectivity (relative difference in fluorescence intensity), we calculate the relative difference in fluorescence intensity between the South and North magnetized substrates, normalized by their sum. The spin selectivity (relative difference in fluorescence intensity) (P) is given by:
\[
P = \frac{I_S - I_N}{I_S + I_N} \times 100
\]

To calculate the uncertainty in spin selectivity (\(P\)), we use the error propagation formula. 
For a function \(P = f(I_S, I_N)\), the propagated uncertainty \(\Delta P\) is:\[\Delta P = \sqrt{\left(\frac{\partial P}{\partial I_S} \Delta I_S\right)^2 + \left(\frac{\partial P}{\partial I_N} \Delta I_N\right)^2}\]
Substituting these into the uncertainty formula gives:\[\Delta P = \sqrt{\left(\frac{2 I_N}{(I_S + I_N)^2} \Delta I_S \times 100\right)^2 + \left(\frac{-2 I_S}{(I_S + I_N)^2} \Delta I_N \times 100\right)^2}\]

We get a spin selectivity (relative difference in fluorescence intensity) of:
P=27±10${\%}$ for the polypeptide, P=5±9${\%}$ for the Phe-Phe peptide, P = 7 ± 9${\%}$ for L-Phe and P = 19 ± 6${\%}$ for D-Phe.

\section{TEM analysis}
\subsection{A-\texorpdfstring{$\beta$}{beta} in Solution VS. on a Substrate}

\begin{figure}[H] 
\centering 
\includegraphics{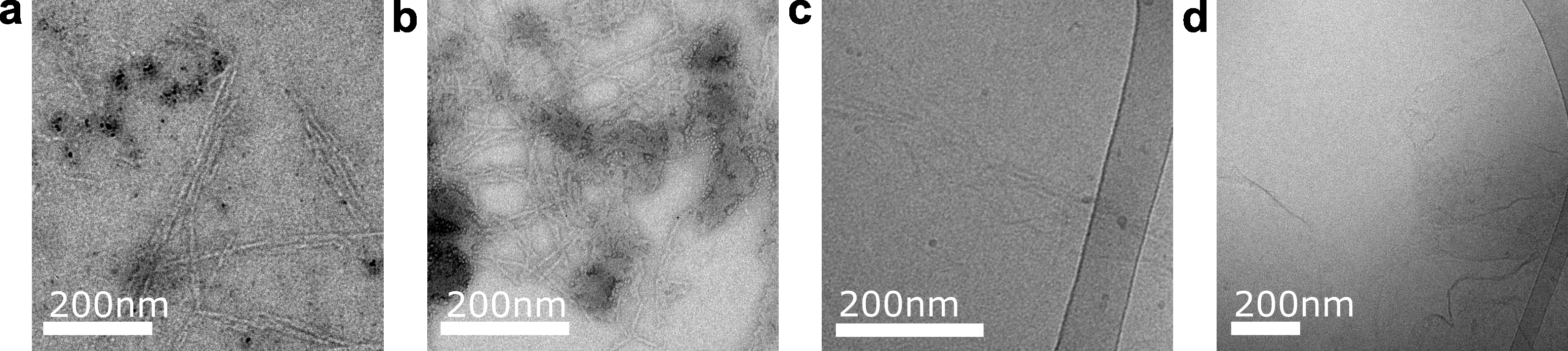} \caption[scheme]{TEM images of A-$\beta$ polypeptide showing \textbf{a} fibrilar aggregates and \textbf{b} fibrilar aggregates formed after 3 days in the solution. Cryo-TEM images of \textbf{c} fibrilar aggregates and \textbf{d} sheets.} 
\label{AbetaTEM} 
\end{figure}
The TEM imaging of negatively stained solution of A-$\beta$ in PBS revealed that the A-$\beta$ polypeptide self-assembled into fibrillar structures and small dark aggregates (\autoref{AbetaTEM}a). The solution was left to age for three days at room temperature, and TEM imaging (\autoref{AbetaTEM}b) revealed similar structures.
Cryo-TEM was employed to observe the difference between the aggregation process in solution and during drying on a substrate. A sample of A-$\beta$ solution in PBS was imaged and showed both fibril structures similar to the structures on the substrates (\autoref{AbetaTEM}c) and wide sheets that were not apparent on the substrate (\autoref{AbetaTEM}d). 

\subsection{TEM of Phe-Phe Peptide}

\begin{figure}[H] 
\centering 
\includegraphics{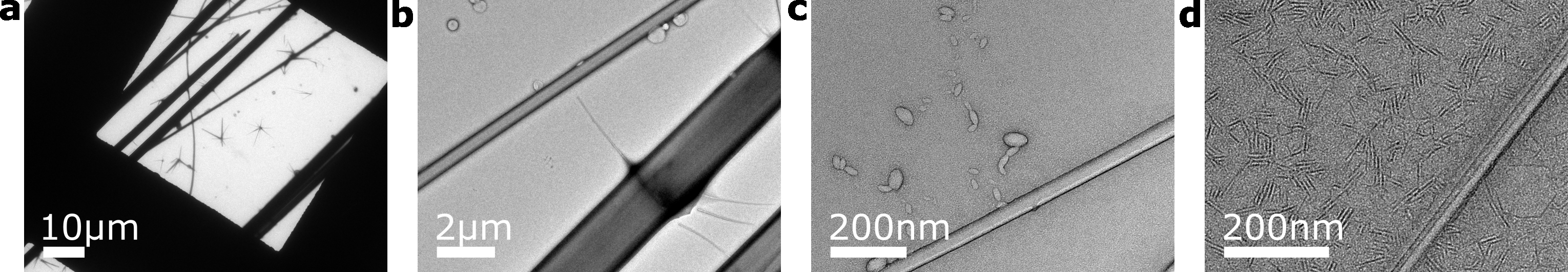} \caption[scheme]{TEM images of the negatively stained sample of PhePhe peptide at different magnifications showing structures that range in size, from longer than the grid hole and 10 $\mu$m diameter (\textbf{a}) to few $\mu$m lond and 5-10 $\mu$m in diameter (\textbf{b-c}). \textbf{d} In addition, very short, uniform structures (20nm long and 2nm diameter) appeared to cover the sample's surface.} 
\label{PhePheTEM} 
\end{figure}

\section{SEM Analysis}

\begin{figure}[H] 
\centering 
\includegraphics{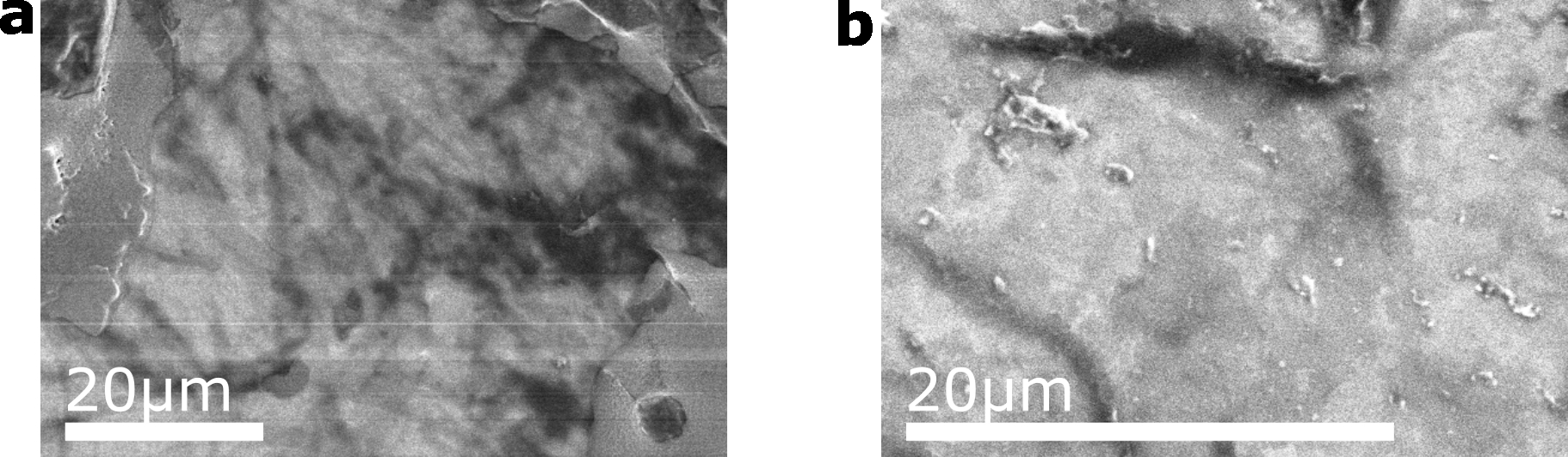} \caption[AbetaSEM]{Secondary electron microscopy (SEM) images of A-$\beta$ polypeptide aggregates. \textbf{a} At 800× magnification, elongated fibrils can be observed bridging two larger aggregates. \textbf{b} At 2000× magnification, darker fibrils are visible, though their fine details are not resolved, along with smaller, brighter aggregates indicating variations in surface topography. Scale bar: 20 $\mu$m.} 
\label{AbetaSEM} 
\end{figure}

\begin{figure}[H] 
\centering 
\includegraphics{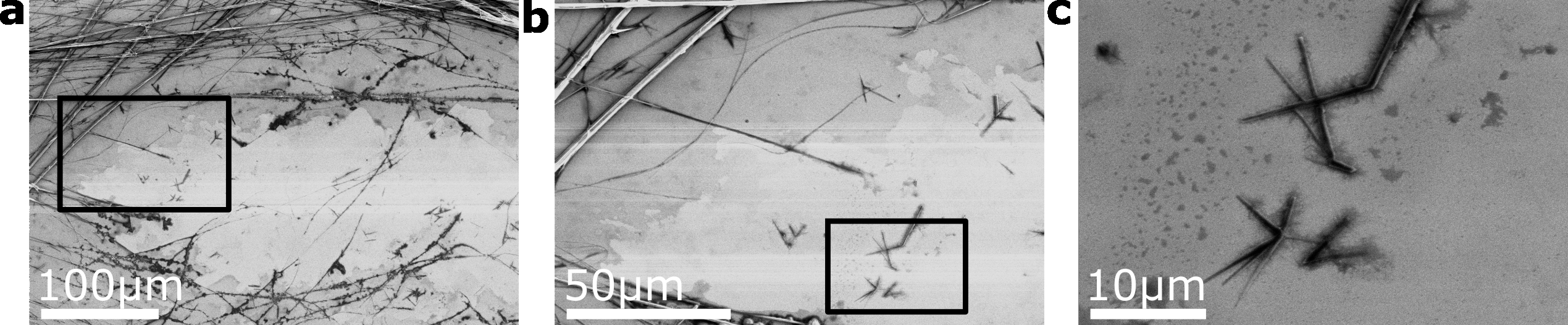} \caption[PhePheSEM]{Secondary electron microscopy (SEM) images of Phe-Phe peptide fibrils. \textbf{a} At 350$\times$ magnification, a variety of fibrils are observed, including long fibrillar structures (>500 $\mu$m) and shorter ones ($\sim$10 $\mu$m), some of which are coated with aggregates. The black rectangle indicates the region enlarged in (b). \textbf{b} At 1000$\times$ magnification, a closer view of the shorter fibrillar structures is shown. The black rectangle marks the area further enlarged in (c). \textbf{c} At 3500$\times$ magnification, the shorter fibrillar structures appear coated with much smaller fibrillar aggregates.} 
\label{PhePheSEM} 
\end{figure}

\begin{figure}[H] 
\centering 
\includegraphics{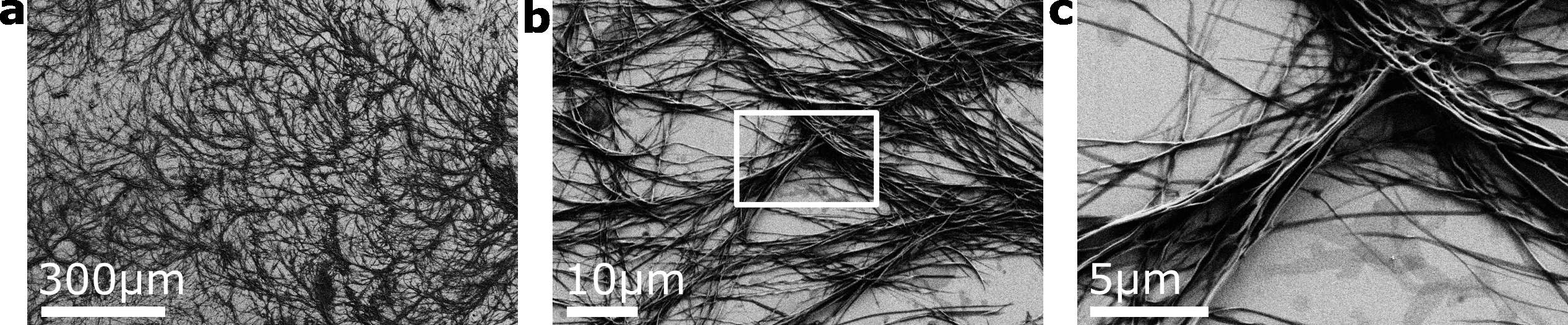} \caption[LPheSEM]{Secondary electron microscopy (SEM) images of L-Phe amino acid fibrils, acquired at 0.1 nA and 2 kV. \textbf{a} At 350$\times$ magnification, fibrillar structures are observed forming bundled arrangements. \textbf{b} At 6000$\times$ magnification, a closer view reveals fibrillar structures with diameters of approximately 500 nm. The white rectangle marks the region enlarged in (c). \textbf{c} At 20,000× magnification, a detailed view of the fibrillar structures is shown.} 
\label{LPheSEM} 
\end{figure}

\begin{figure}[H] 
\centering 
\includegraphics{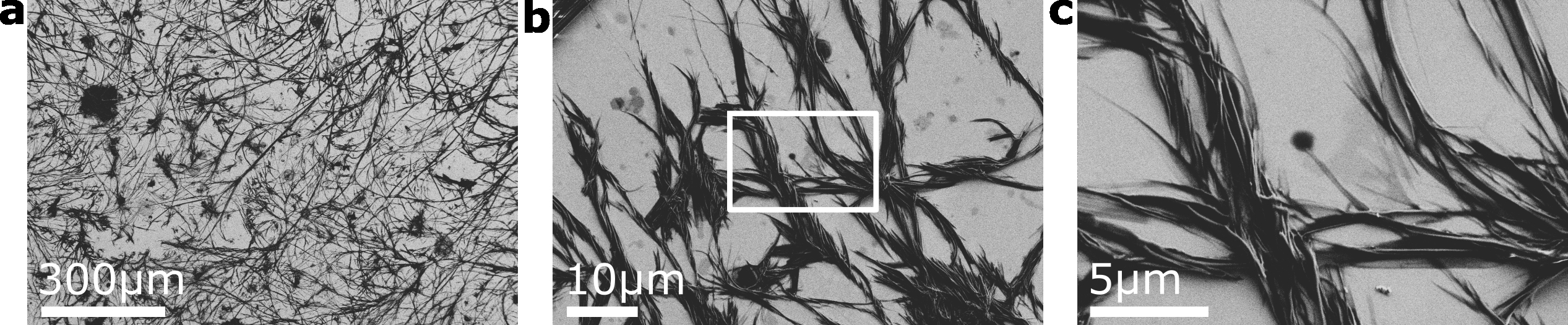} \caption[DPheSEM]{Secondary electron microscopy (SEM) images of D-Phe amino acid fibrils, acquired at 0.1 nA and 2 kV. \textbf{a} At 350$\times$ magnification, fibrillar structures are observed forming bundled arrangements. \textbf{b} At 6000$\times$ magnification, a closer view reveals fibrillar structures with diameters of approximately 500 nm. The white rectangle marks the region enlarged in (c). \textbf{c} At 20,000× magnification, a detailed view of the fibrillar structures is shown. The fibrils seem shorter han the L-chirality fibrils.} 
\label{DPheSEM} 
\end{figure}

\section{Confocal Images Analysis}
\subsection{A-\texorpdfstring{$\beta$}{beta} polypeptide and Phe-Phe Peptide}
Fluorescence intensity and its error are calculated by acquiring three confocal fluorescence images for each of the two samples. For each image, a Z-projection is performed using the ImageJ program to obtain an average-intensity image, from which the mean fluorescence intensity across the entire image is computed. These mean intensities are then averaged over all three images to determine the final average fluorescence signal (in arbitrary units, au) for each sample. The error is represented by the standard error of the mean (SEM) calculated from the fluorescence intensities of the different images.
\subsection{Phe Amino Acid}
Fluorescence intensity and its error are calculated by acquiring fluorescence images from the magnetic or non-magnetic area of the sample. For each image, a Z-projection is performed using the ImageJ program to obtain an average intensity image, and a region of interest (ROI) is selected to ensure sufficient distance from the edge of the drop, avoiding the effects of the coffee ring. The ROI is then divided into four parts, and the mean intensity is calculated for each. These mean intensities are averaged to determine the final average fluorescence signal (in arbitrary units, au) for each sample. The error is represented by the standard error of the mean (SEM) calculated from the fluorescence intensities of the four areas.
\subsection{Image Files}
Full image files are available at https://doi.org/10.5281/zenodo.14925254

\subsection{Baseline ThT Fluorescence}

\begin{figure}[H] 
\centering 
\includegraphics{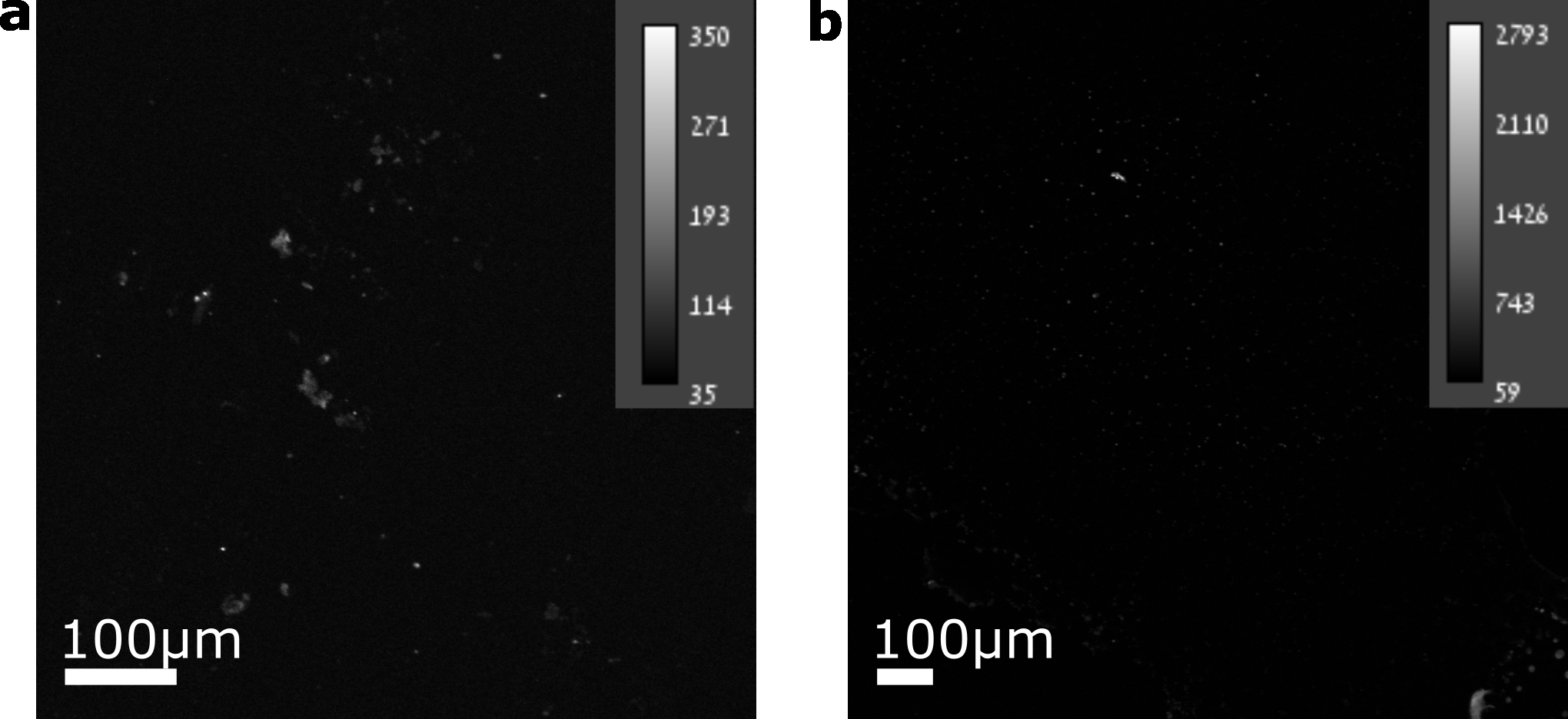} \caption[ThTimageSI]{Z-projection images of ThT fluorescence intensity obtained using ImageJ. \textbf{a} 40 $\mu$M ThT in PBS, representing the baseline fluorescence intensity for A-$\beta$ polypeptide samples, with an average intensity of 48. \textbf{b} 10 $\mu$M ThT in water, representing the baseline fluorescence intensity for Phe-Phe peptide and L-Phe amino acid samples, with an average intensity of 82.} 
\label{ThTimageSI} 
\end{figure}
\subsection{ThT Fluorescence Spectrum}

\begin{figure}[H] 
\centering 
\includegraphics{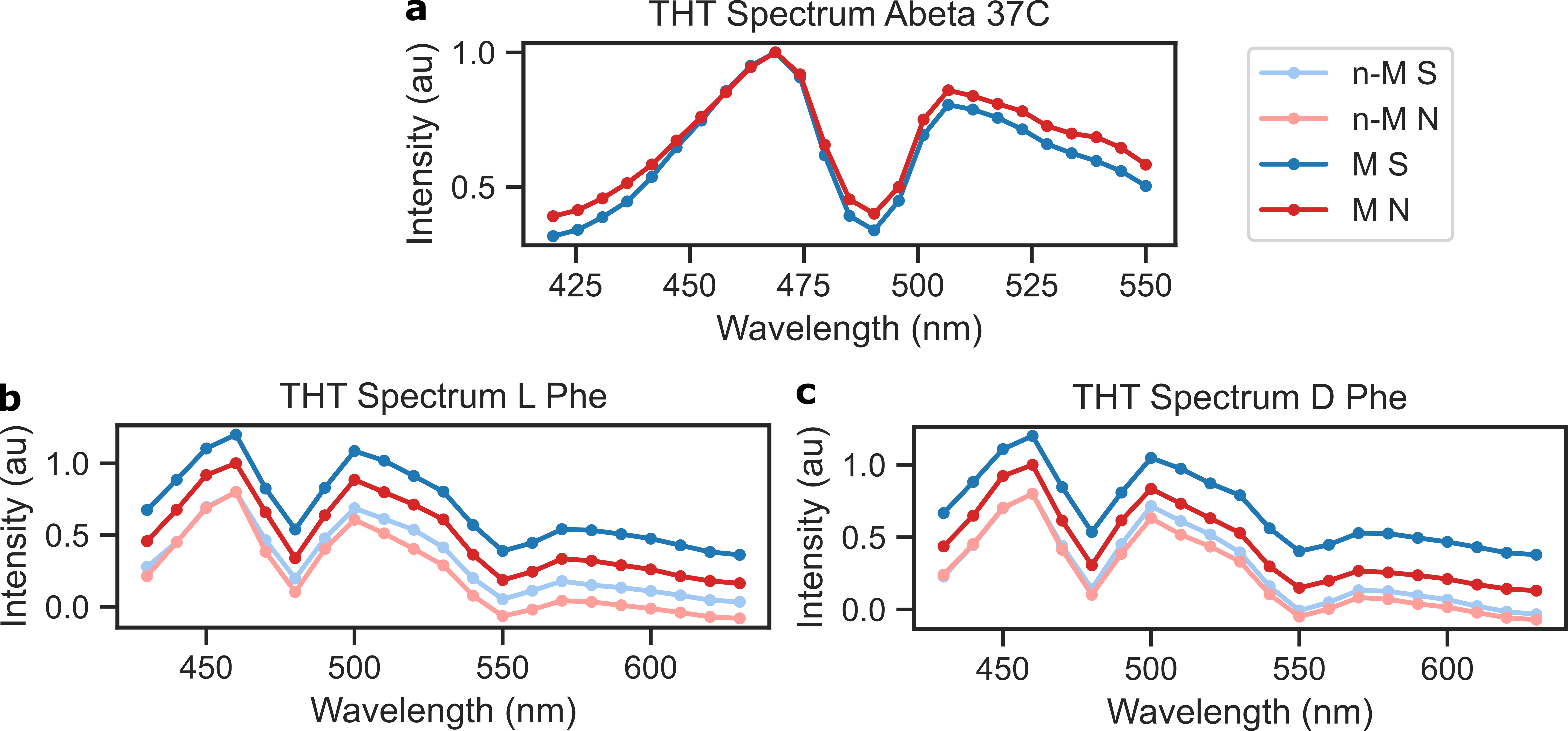} \caption[ThTspectraSI]{Fluorescence spectra of ThT for the samples. The spectra remain unchanged between different samples and magnetization orientations. The South-oriented spectra are shown in blue, North-oriented spectra in red, and nonmagnetic sides in light blue/red. \textbf{a} Fluorescence spectra of the A-$\beta$ sample. \textbf{b} Fluorescence spectra of the L-Phe sample. \textbf{c} Fluorescence spectra of the D-Phe sample.} 
\label{ThTspectraSI} 
\end{figure}

\section{FTIR Analysis}
In our ATR-FTIR analysis, we focused on the Amide region (1700–1500 cm$^{-1}$), which strongly correlates with amyloid fibrils, as reported by Sarroukh et al. \cite{sarroukh2013atr}. Here, we describe the distinct peaks observed in different samples and their corresponding conformations.

In the Amide I range (1700–1600 cm$^{-1}$), several $\beta$-sheet signatures were identified, consistent with previous findings \cite{sarroukh2013atr}. The A-$\beta$ sample exhibits a strong peak at 1630 cm$^{-1}$, a characteristic marker of amyloid fibrils. The dipeptide spectrum features a peak at 1685 cm$^{-1}$, signifying the presence of anti-parallel $\beta$-sheets (APB). The presence of APB suggests that oligomers coexist with mature fibril structures in the sample. Similarly, in the A-$\beta$ spectrum, a small shoulder appears in the same region, indicating the presence of oligomers in this sample as well. The single amino acid samples show a peak at 1640 cm$^{-1}$, which is indicative of parallel $\beta$-sheets (PB). 

The distinct peak at 1605 cm$^{-1}$ is attributed to a C-C ring vibration, which is notably absent in the A-$\beta$ polypeptide spectrum. This observation is consistent with findings in other polypeptides containing phenylalanine \cite{barth2000infrared}, where the side chain vibrations do not prominently appear in FTIR spectra.

A peak in the Amide II region (1600–1500 cm$^{-1}$) exhibits a shift between the North and South directions of magnetization. This 30 cm$^{-1}$ shift is observed in both the A-$\beta$ polypeptide (from 1530 to 1560 cm$^{-1}$) and the dipeptide Phe-Phe (from 1550 to 1580 cm$^{-1}$) but is absent in the single amino acid. This suggests a decrease in anti-parallel $\beta$-sheets (APBs) as aggregation progresses. \cite{sarroukh2011transformation}. It is likely that there are both oligomers and mature fibrils in each sample.

\section{Charge Termini Peptide}
\begin{figure}[H] 
\centering 
\includegraphics{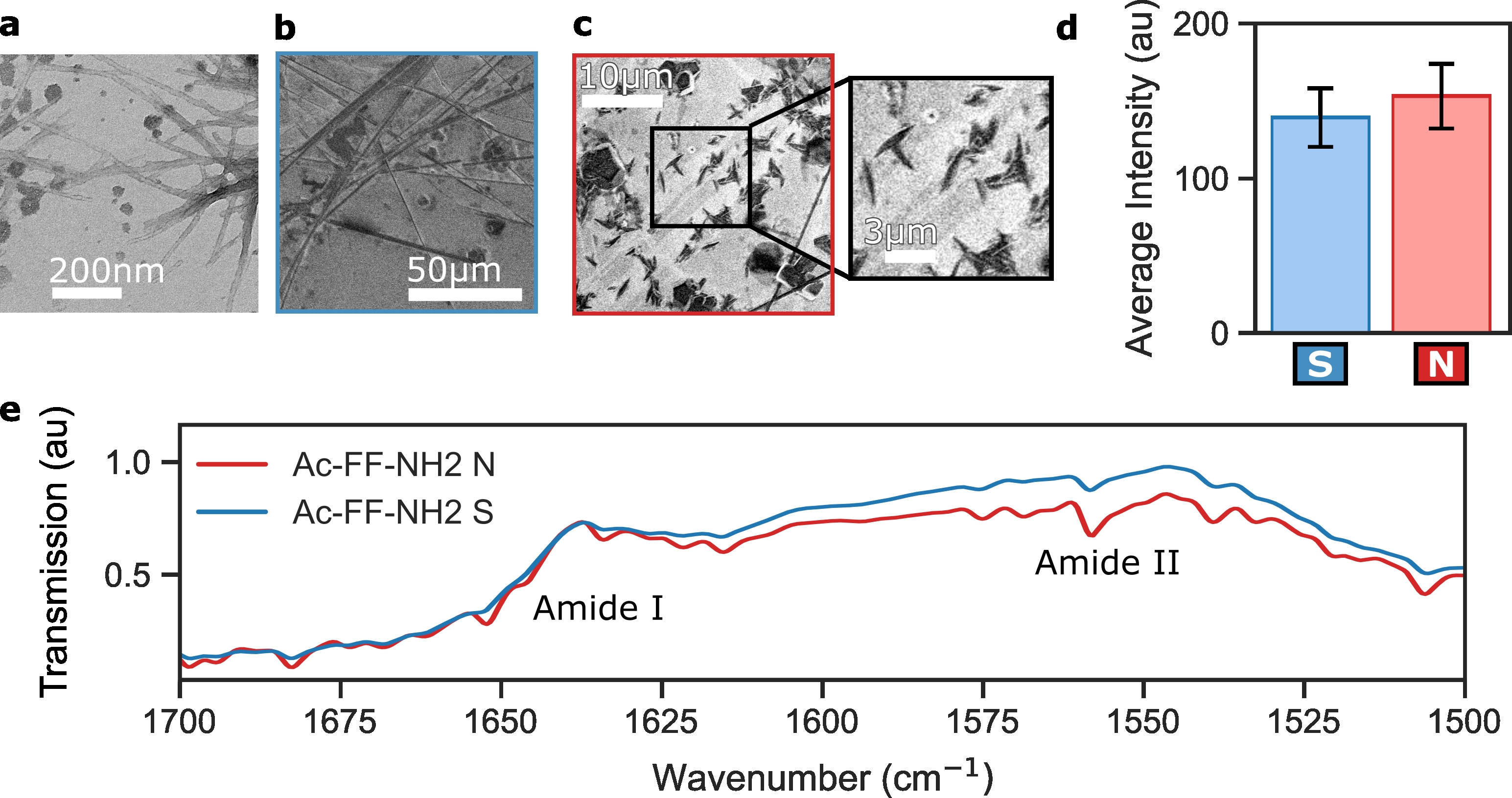} \caption[AcPhe2NH2]{\textbf{a} TEM image showing Ac-Phe–Phe-NH$_2$ formed fibrillar aggregates. \textbf{b} SEM image of Ac-Phe–Phe-NH$_2$ fibrils formed on a South-magnetized substrate. \textbf{c} SEM image of Phe-Phe fibrils formed on a North-magnetized substrate. \textbf{d} Average fluorescence intensity of ThT dyed fibrils.\textbf{e} FTIR spectra of Ac-Phe–Phe-NH$_2$ fibrils formed on South- or North-magnetized substrate.} 
\label{AcPhe2NH2} 
\end{figure}

To exclude the potential contribution of electrostatic interactions in the assembly process, we studied a modified, non-charged peptide analogue, Ac-Phe–Phe-NH$_2$, in which the N-terminal amine was acetylated and the C-terminal carboxyl was amidated. This modification reduces the peptide's dipole moment compared to the native form.\cite{reches2005self} TEM imaging presented in \autoref{AcPhe2NH2}a confirmed that the analogue self-assembles into fibrillar structures (diameter 25nm and length 300nm). SEM images on North- and South- magnetized substrates are presented in \autoref{AcPhe2NH2}b,c. Significantly shorter fibrils were observed on the North side ($>$200$\mu$m vs. $\sim$3$\mu$m, similar to the behavior observed for the natural peptide. In addition, the assembly was more uniform on the South magnetized substrate. Fluorescence intensity measurements using a ThT dye (\autoref{AcPhe2NH2}d) and FTIR spectroscopy (\autoref{AcPhe2NH2}e) showed no significant differences between fibrils formed on North- versus South-magnetized substrates. FTIR spectroscopy revealed an Amide I peak at 1640 cm$^{-1}$ and an Amide II peak at 1545 cm$^{-1}$ corresponding to the known ${\beta}$-sheet signatures as described in Reches et al. \cite{reches2005self}

Our results show that while the peptide analogue forms similar fibrillar morphologies, its structural behavior differs from the native peptide. This suggests that electrostatic interactions may play a role in the assembly process and that these interactions could be influenced by spin effects.
\end{document}